\documentclass[12pt]{iopart}

\usepackage{iopams}
\expandafter\let\csname equation*\endcsname\relax

\expandafter\let\csname endequation*\endcsname\relax

\usepackage{amsmath}
\usepackage{graphicx}% Include figure files
\usepackage[breaklinks=true,colorlinks=true,linkcolor=blue,urlcolor=blue,citecolor=blue]{hyperref}
\usepackage{ulem}
\usepackage{silence}
\usepackage{verbatim}
\usepackage{bm}% bold math
\usepackage{lineno}

\newcommand{\beq}{\begin{equation}}
\newcommand{\eeq}{\end{equation}}

\begin{document}

\title[Spacing ratio statistics of multiplex directed networks]{Spacing ratio statistics of multiplex directed networks}

\author{Tanu Raghav, and Sarika Jalan}
\address{Complex Systems Lab, Indian Institute of Technology Indore - Simrol, Indore - 453552, India}

%\ead{*Corresponding Author:sarika@iiti.ac.in}
\vspace{10pt}
%\begin{indented}
%\item[\today]%October 2020
%\end{indented}

\begin{abstract}
Eigenvalues statistics of various many-body systems have been widely studied using the nearest neighbor spacing distribution under the random matrix theory framework. Here, we  numerically analyze eigenvalue ratio statistics of multiplex networks consisting of directed Erd\H{o}s-R\'{e}nyi random networks layers represented as, first, weighted non-Hermitian random matrices and then weighted Hermitian random matrices. We report that the multiplexing strength rules the behavior of average spacing ratio statistics for multiplexing networks represented by the non-Hermitian and Hermitian matrices, respectively. Additionally, for both these representations of the directed multiplex networks, the multiplexing strength appears as a guiding parameter for the eigenvector delocalization of the entire system. These results could be important for driving dynamical processes in several real-world multilayer systems, particularly, understanding the significance of multiplexing in comprehending network properties.
\end{abstract}

\vspace{2pc}
\noindent{\it Keywords}: Eigenvalues, RMT, multiplex network
%\submitto{\NJP}
%\maketitle
\section{Introduction}
%In $1950s$, several many-body complex quantum systems were analyzed using eigenvalues and eigenfunctions statistics under random matrix theory (RMT) framework. 
Random matrix theory (RMT) has been successful in finding universality among spectral correlations of enormous class of many-body complex systems. Under this theory, instead of investigating a system’s Hamiltonian, an ensemble of random Hamiltonian matrices is investigated to outline the system specific properties \cite{Guhr}. RMT has been useful in many areas of physics and mathematics ranging from condensed matter to economical financial markets by successfully describing properties of spectral fluctuation of complex nuclei, atoms, and complex molecules \cite{app_RMT}. One of the most commonly used measures in RMT is nearest neighbor spacing distribution (NNSD) \cite{Mehta}. NNSD corresponds to the probability density $P(s)$ of the differences ($s_i$) between two consecutive levels, say $\epsilon_{i+1}-\epsilon_i$ with i=$1, 2, ..., N$. For quantum systems with regular classical counterparts, NNSD coincides with the Poisson distribution $P(s)= \exp(-s)$, whereas for the Gaussian orthogonal ensemble (GOE) and Gaussian unitary ensemble (GUE), NNSD can be well approximated by the Wigner distribution \cite{Mehta} 
$P(s)_{GOE} = \frac{\pi}{2}s\exp\left(- \frac{{\pi} s^2}{4}\right)$ and $P(s)_{GUE} = \frac{32}{\pi^2}s^2\exp\left(- \frac{{4} s^2}{\pi}\right)$, respectively.

\begin{figure}[t!]
	\begin{center}
	\includegraphics[height=7cm,width=10cm]{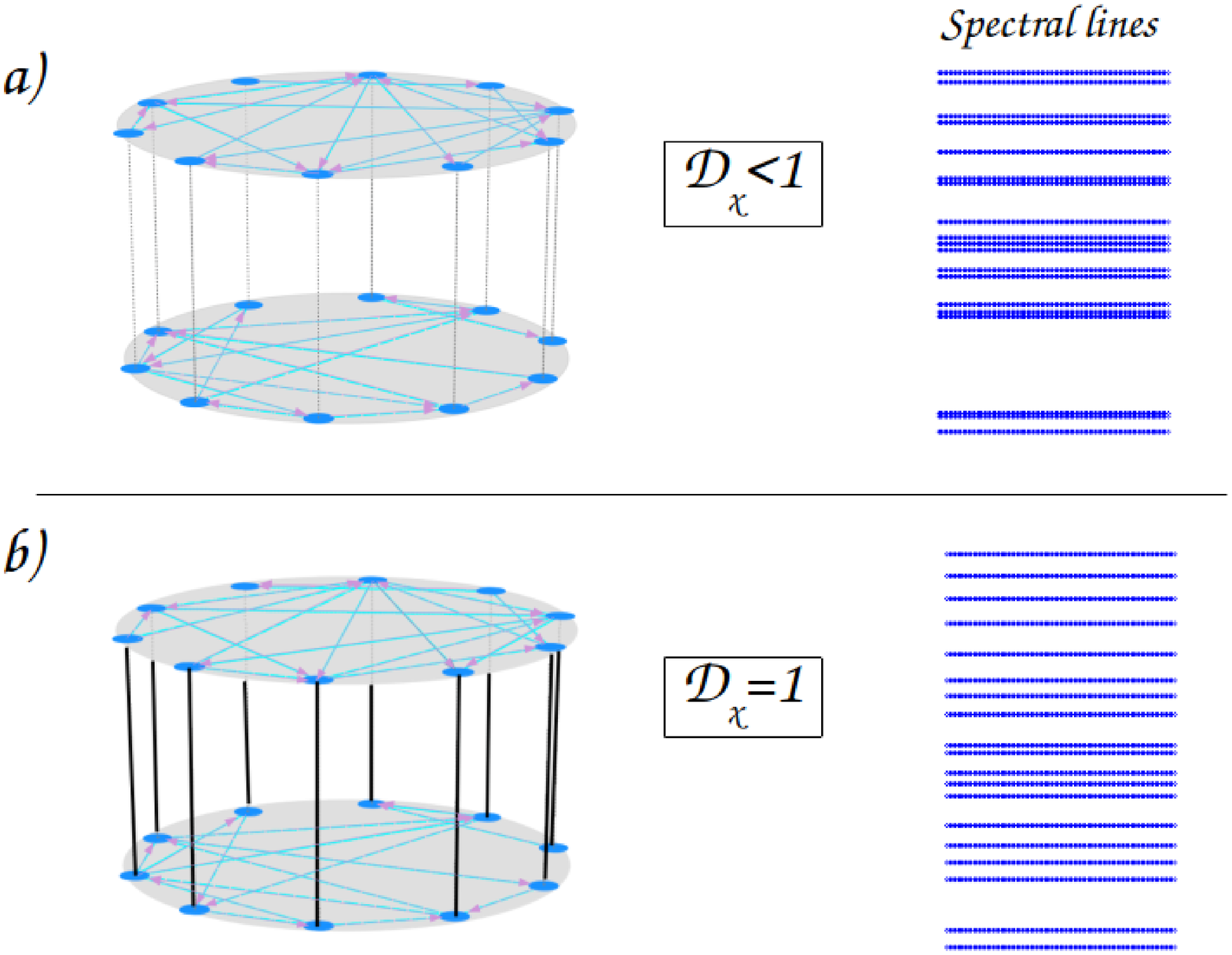}\\
	\vspace{-0.5cm}
	\caption{{\bf Schematic representation of bi-layer multiplex network}. Intra-layer connections are configured as directed edges whereas inter-layer connections are bi-directional. (a) Multiplex network and spectral lines for weak multiplexing strength ($D_x<1$), (b) strong multiplexing ($D_x=1$).}
	\end{center}
	\label{figure1}
\end{figure}

To calculate NNSD, one has to eliminate the impact of level density by a process called unfolding so that the mean level spacing becomes one \cite{unf}. However, due to the unavailability of an analytical form of the mean level density, the spectrum is unfolded numerically. Unfolding the spectra can be ambiguous as it requires polynomial curve fitting by choosing the fitting parameter arbitrarily \cite{unf1}. To avoid this perplexity, a measure has been introduced called the eigenvalue spacing ratio, defined as the ratio between the nearest and next nearest neighbor spacing of the eigenvalues \cite{rat_int}. Since average mean spacing becomes insignificant in calculating spacing ratios, no unfolding is required. Over the years, the spacing ratio has been widely used to study the spectral statistics of several many-body systems, including complex networks \cite{rat_app, comp_net, sl_rat}.

Furthermore, complex systems are widely studied using network theory by recognizing that several real-world systems around us comprise of interacting units or nodes \cite{Rev_N}. In recent years, network theory has been successful in providing insights into structure and dynamics of physical, biological, social, and several other natural systems \cite{Rev}. Recently, the realization that various types of interactions may exist between the same set of nodes has paved the way for multilayer networks which not only incorporate different types of connectivity in individual layers but also provide a representation of interconnected systems \cite{MN1}. These interconnected systems conceptualized as multilayer networks give new insights into several structural and dynamical properties compared to their monolayer counterparts \cite{MN2, MN3}. An elemental feature of representing multilayer networks is assessing the interconnectivity between various types of connections. Investigating such interconnectivity or inter-layer connections is essential for studying the role of dynamical switching between layers in a multiplex system. These inter-layer connections weights, referred to as multiplexing strength, provide a quantitative measure of impacts of dynamical and structural properties of one layer on those of the another layer \cite{MN4, MN5}. Various studies on multilayer networks have established the emerging aspects of the interplay between the intra and inter-layer connections of the associated network \cite{MN6, MN7}. In recent times, a new emerging paradigm in network science encapsulating higher order interactions has gained an immense population. The studies associated with higher order interactions under the framework of hypergraphs and simplicial complexes have unveiled interesting phenomena about complex networks \cite{Majhi}.

Next, on a fundamental level, interactions are asymmetrical in nature for a large class of networks corresponding to real-world complex systems. In a network, a connection between two nodes often has a particular direction mathematically represented by directed networks and corresponding asymmetric adjacency matrices. Well-known examples include the food web network \cite{food-web}, gene regulatory network \cite{gene}, citation \cite{cit}, and world wide web networks \cite{www}. Also, the spectra of adjacency matrix of a network \cite{graph} relate to various structural properties \cite{str_prop1, str_prop2, dismantle}, as well as dynamical processes on the underlying networks \cite{dyn_prop2, nine}. Nevertheless, spectral properties of directed networks \cite{SJ_dir, bin_ye1, c_li} have not been investigated in as much detail as for undirected networks \cite{Farkas, cam}. Regardless of the significance of directed complex networks, spectral properties of directed networks are yet to be explored more to understand the system's dynamical processes and their relation with the structural properties. Because of the complex eigenvalues of the asymmetrical adjacency matrix, the analytical approach can be challenging to analyze spectra of directed networks. Investigations of complex eigenvalues corresponding to directed networks and several other many-body systems have gained momentum in recent years. For example, J. Baron analyzed the spectra of directed complex networks owning heterogeneity in the network, and deduced effect of heterogeneity on the stability of corresponding system \cite{arx}. Recently, Metz and Neri analytically obtained an exact expressions for the inverse participation ratio by relating it with the structural properties of the directed networks. They showed that as network connectivity increases, the largest eigenvalue and the eigenvalue at the boundary of the bulk undergo from a localized to a delocalized regime \cite{Neri}.

%%%%%%%%%%%%%%%%%%%%%%%%%%%%%%%%%%%%%%%%%%%%%%%%%%%%%%%%%%%%%%%%%%%%%%%%%%%%%%%%%%%%%%%%
\begin{figure*}[b!]
	\begin{center}
	\includegraphics[height=3.5cm,width=9cm]{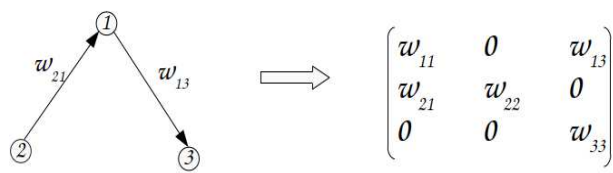}\\
	\vspace{-0.5cm}
	\caption{Illustration of a network and corresponding adjacency matrix represented by the diluted real Ginibre ensemble ($A_{dRGE}$)}.
	\label{fig.01}
	\end{center}
\end{figure*}
%%%%%%%%%%%%%%%%%%%%%%%%%%%%%%%%%%%%%%%%%%%%%%%%%%%%%%%%%%%%%%%%%%%%%%%%%%%%%%%%%%%%%%%%

Further, T. Peron and colleagues analyzed the spectra of directed single-layer networks using the eigenvalue spacing ratio ($r$). By doing so, the authors set out that the spectral properties can be relevant to dynamical processes on the directed networks, and deduced the network parameters that can control dynamical transitions in the corresponding system \cite{sl_rat}. In this paper, we extend the spacing ratio analysis of directed single-layer networks to multiplex networks consisting of two layers. One of the questions which we focus on is that for multiplexing networks what could be governing parameters for any such transition which is observed in single-layer networks. Both the layers are constructed using directed Erd\H{o}s-R\'{e}nyi (ER) random networks, represented by (i) weighted non-Hermitian random adjacency matrix and (ii) Hermitian random adjacency matrix. Notably, we use the average spacing ratio $\langle r\rangle$ as a measure of complexity in the system as a function of different structural parameters of the corresponding networks. Later, $\langle r\rangle$ is used to identify the delocalization transition of directed multiplex networks. We show that multiplexing strength not only can change the minimum probability required to commence the $\langle r\rangle$ transition, but also can suppresses the transition. 

The paper is organized as follows. Section.~\ref{s2} comprises the model definition and construction along with the definition of technique employed in this paper i.e., eigenvalue spacing ratio for real and complex eigenvalues. Section.~\ref{s31} and section.~\ref{s32} discuss the results on the interplay of the ratio statistics and various structural parameters for Non-Hermitian and Hermitian representation of multiplex network. Section.~\ref{s4} associates the localization-delocalization transition using Shannon entropy with the average ratio transition from the Poisson ensemble to GUE/RGE where RGE stands for real Ginibre ensemble. Section.~\ref{s5} concludes the paper with future applications and directions.

\section{\label{s2}Models and quantities}
A multiplex network can be defined as $G=(V, E),$ where $V=V_1 \cup \cdots \cup V_m$, $V_i$ being the set of nodes in the layer $i$, $E=\{(a, b) | a \in V_i, b \in V_j\}$. If $i=j$, the edge $(a, b)$ is an intra-layer connection; that is, it is on the nodes of the same layer. If $i \neq j$, the edge $(a, b)$ is an inter-layer edge between the nodes of layer $i$ and layer $j$. There exists a special type of multilayer network referred as multiplex network, wherein each layer there exists same set of nodes, that is, $V_1=\dots=V_m$. In such networks, inter-layer connections connect each node of one layer to its mirror node in the other layers. Here, we consider bi-layer multiplex networks, with each layer having $N$ nodes. The adjacency matrix $A$ for such a $2$-layer multiplex network can be written as
\begin{eqnarray} 
\label{eq1}
A =  \left(
\begin{array}{cc}
\ A^1  & D_{x}I \\ 
D_{x}I & \ A^2 \\
\end{array}
\right)
\end{eqnarray}
where $ A^{1} $ and $ A^{2} $ are the adjacency matrices corresponding to the layer $1$ and layer $2$, respectively, and $D_x$ is the multiplexing or inter-layer strength. Degree of the $i$-th node in any layer $m$ ($m=1, 2$) is given by $ {k_{i}}^{m} $ = $\sum_{j=1}^{N}$ ${A_{ij}}^m + D_x$ where ${A_{ij}}^m$ is the $ij$-th entry of $A^{m}$ matrix and $D_x$ corresponds to the inter-layer connection. Average degree of a network in layer $m$ is then $\langle k\rangle^m$=$\sum_{i=1}^{N}\frac{k_{i}^{m}}{N}$. The eigenvalues of an adjacency matrix $A$ of a multiplex network are denoted by $ \lambda_{i} $, $i = 1, \dots, 2N$, where $2N$ is the size of the multiplex network and $\lambda_{1} \le \lambda_{2} \le \lambda_{3} \le \dots \leq \lambda_{2N-1} < \lambda_{2N}$.

%%%%%%%%%%%%%%%%%%%%%%%%%%%%%%%%%%%%%%%%%%%%%%%%%%%%%%%%%%%%%%%%%%%%%%%%%%%%%%%%%%%%%%%%
\begin{figure*}[t!]
	\begin{center}
	\includegraphics[height=4cm,width=9cm]{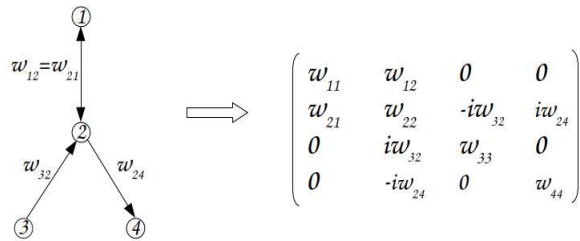}\\
	\vspace{-0.5cm}
	\caption{An example of a network describing magnetic adjacency matrix ($A_M$).}
	\label{fig.02}
	\end{center}
\end{figure*}
%%%%%%%%%%%%%%%%%%%%%%%%%%%%%%%%%%%%%%%%%%%%%%%%%%%%%%%%%%%%%%%%%%%%%%%%%%%%%%%%%%%%%%%%

\subsection{ Multiplex network with random weighted non-Hermitian matrix}

In a multiplex network, we consider ER random networks for both the layers to generate a directed random network \cite{R1}. The ER network of size $N$ is constructed by connecting each pair of the nodes by a directed edge with a probability $p$. Next, the edges are weighted with statistically independent random variables $w$, drawn from a normal distribution (mean$=0$ and variance$=1$). If there exists a directed edge from $i$ to $j$, we assign a weight $w_{ij}$ and for $i=j$, the weight is $w_{ii}$ (Fig.~\ref{fig.01}). We denote the weighted adjacency matrix of both the layers by $A_{dRGE}$ and define $a_{ij}$ (for layers 1, 2) as  
%%%%%%%%%%%%%%%%%%%%%%%%%%%%%%%%%%%%%%%%%%%%%%%%%%%%%%%
\begin{equation}\label{eq2}
a_{ij}=\left\{
\begin{array}{ll}
w_{ii} & \mbox{if $i=j$}, \\
w_{ij} & \mbox{if $i\rightarrow j$}, \\
0 & \mbox{otherwise}.
\end{array}
\right.
\end{equation}
%%%%%%%%%%%%%%%%%%%%%%%%%%%%%%%%%%%%%%%%%%%%%%%%%%%%%%%%%

Because of the directed nature of the network ($w_{ij} \neq w_{ji}$) the matrix $A_{dRGE}$ can be non-Hermitian. Such matrices are identified and have been defined as a diluted version of the real Ginibre ensemble (dRGE) \cite{sl_rat}. Real Ginibre ensemble (RGE) comprises of $n\times n$ random matrices formed from $i.i.d$ standard Gaussian variables. Note that, for $p=1$ (complete network), $A_{dRGE}$ is drawn from the RGE distribution.  Also, for $p=0$, i.e., a completely disconnected network, $A_{dRGE}$ becomes a diagonal random matrix and follows Poisson ensemble (PE) statistics. Therefore, for the adjacency matrix of multiplex network represented by $A_{NH}$, the block matrices $A^1$ and $A^2$ are independently distributed as $A_{dRGE}$ representing layer $1$ and layer $2$, respectively, whereas off-diagonal block matrices ($D_xI$) correspond to bi-directional inter-layer connections where $I$ is the identity matrix connecting mirror nodes of layers $1$ and $2$.

\subsection{Multiplex network with random weighted Hermitian matrix}
 To study the Hermitian representation of the adjacency matrix of directed random graphs, we again consider ER random model for both the layers. Next, in the extracted adjacency matrix, the edges are weighted with statistically independent random variable $w$ (Fig.~\ref{fig.02}) drawn from normal distribution (zero mean and variance one). We denote the weighted adjacency matrix as $A_M$ for the layers $1$ and $2$, and define $A_M=(a_{ij})$ as  
\begin{equation}\label{eq3}
a_{ij}=\left\{
\begin{array}{ll}
w_{ii} & \mbox{if $i=j$}, \\
w_{ij} & \mbox{if $i\leftrightarrow j$}, \\
\iota w_{ij} & \mbox{if $i\rightarrow j$}, \\
-\iota w_{ij} & \mbox{if $j\rightarrow i$}, \\
0 & \mbox{otherwise}.
\end{array}
\right.
\end{equation} 

Above defined method has been recently introduced in References \cite{mix1, mix2}. Lately, the adjacency operator defined in Refs. \cite{mix1, mix2} has been used in various studies such as semi-directed graphs \cite{sd}, cluster hypergraphs of semi-directed graphs \cite{hsd}, quantum walk \cite{qw} etc.
Note that $A_H$ is constructed such that $[A_H]^*_{uv}$ = $[A_H]_{vu}$. Here, as $p$ is increased, a transition is observed from the PE ($p$=$0$) to fully connected real symmetric matrices ($p$=$1$). At $p$=$1$, the ensemble is similar to the Gaussian orthogonal ensemble (GOE) of random matrix theory (RMT), with the difference that; in the GOE, the diagonal matrix elements have twice the variance as the off-diagonal ones. Such a matrix has been defined as a magnetic adjacency matrix ($A_M$) \cite{sl_rat}. Here, in the adjacency matrix of a multiplex network represented by $A_H$, $A_1$ and $A_2$ are represented by $A_M$, constituting both layers of the multiplex network.

\subsection{Measures used for spectral properties}{\label{s23}
To study the spectral properties of multiplex networks with both layers consisting of directed random networks, we consider eigenvalue ratio statistics. The eigenvalue ratio statistics has been extensively used to characterize the localization to de-localization transition and, thus, Poisson to GOE \cite{rat_loc}. In the case of the non-Hermitian adjacency matrix, complex eigenvalues ($\lambda_\mathbb{C}$) are obtained, and thus, complex spacing ratio $r_\mathbb{C}$ is computed in the following manner \cite{rat3}. $r_\mathbb{C}$ is defined as the ratio ($i$-th) between the distance of the nearest neighbor eigenvalue to the distance of the next to nearest neighbor eigenvalue,
\begin{equation}\label{eq4}
    r^i_\mathbb{C}= \frac{\lvert \lambda^{NN}_i - \lambda_i \rvert}{\lvert \lambda^{NNN}_i - \lambda_i \rvert}
\end{equation}
where $\lambda^{NN}_i$ and $\lambda^{NNN}_i$ are nearest and next nearest neighbor respectively, of $\lambda_i$. However, for real eigenvalues ($\lambda_\mathbb{R}$), in case of the Hermitian adjacency matrix, the $i$-th ratio

\begin{equation}\label{eq5}
    r^i_\mathbb{R}= \frac{min(\lambda_{i+1} - \lambda_i, \lambda_i - \lambda_{i-1}) }{max(\lambda_{i+1} - \lambda_i , \lambda_i - \lambda_{i-1})}
\end{equation}
Here, $r_\mathbb{R}$ and $r_\mathbb{C} \in [0,1]$. 
For $N=3$, the probability distribution function of $r_\mathbb{R}$ for the PE and GUE \cite{rat2} is given by,
\begin{equation}\label{eq6}
    P_{PE}(r_\mathbb{R}) = \frac{2}{(1+r_\mathbb{R})^2}
\end{equation}
\begin{equation}\label{eq7}
    {P_{GUE}}(r_\mathbb{R}) = \frac{81\sqrt{3}}{2\pi}\frac{(r_\mathbb{R}+r_\mathbb{R}^2)^2}{(1+r_\mathbb{R}+r_\mathbb{R}^2)^4} \ ,
\end{equation}
GUE consists of random $n \times n$ Hermitian matrices whose complex entries (both real and imaginary) are $i.i.d$ Gaussian variables. For GOE \cite{rat2},
\begin{equation}\label{eq8}
{P_{GOE}}(r_\mathbb{R}) = \frac{27}{4}\frac{r_\mathbb{R}+r_\mathbb{R}^2}{(1+r_\mathbb{R}+r_\mathbb{R}^2)^{5/2}} \ ,
\end{equation}

Eq.~(\ref{eq6}) to (\ref{eq8}) provide good approximations for $N\gg3$ also. In the case of $r_\mathbb{C}$, $P_{RGE}(r_\mathbb{C})$ is not known. To validate our results for $P(r_\mathbb{C})$, we refer to the work \cite{sl_rat}. However, recently, Dusa and Wettig derived an approximate expression of the probability distribution $P_{RGE}(r_\mathbb{C})$ for the GUE \cite{gini_new}.

%%%%%%%%%%%%%%%%%%%%%%%%%%%%%%%%%%%%%%%%%%%%%%%%%%%%%%%%
\begin{figure*}[b]
\begin{center}
\includegraphics[height=6.5cm,width=8cm]{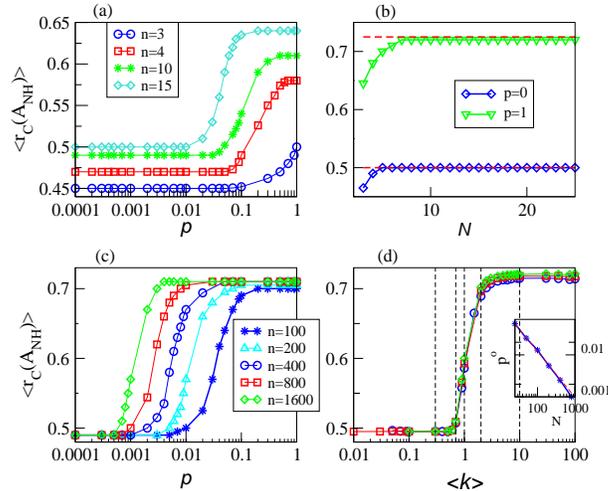}\\
\vspace{-0.5cm}
\caption{(a)-(c) Average spacing ratio $r_\mathbb{C}$ as a function of the connection probability $p$ in both layers of a multiplex network represented by the diluted real Ginibre ensemble for different network sizes $N$. Horizontal dashed lines (b) for $p$=$0$ and $p$=$1$ are marked at $r_\mathbb{C}$=$0.5$ and $0.73$, respectively, (c) marked at $r_\mathbb{C}$=$0.60$. (d) $r_\mathbb{C}$ as a function of $\langle k\rangle$ for different network sizes as in (c). In the inset, $p_o$ as a function of $N$ fitted by Eq.~(\ref{eq9}). Vertical dashed lines indicate $\langle k\rangle$=$0.3$, $0.7$, $1$, $2$ and $10$. Each data point is averaged over $10^6/2N$ random realizations. Increase in network size initiates the transition for lower $p$ whereas once the $\langle k\rangle$ is fixed, $r_\mathbb{C}$ is also fixed. }
\label{fig.1}
\end{center}
\end{figure*}

\begin{figure*}[t!]
	\begin{center}
	\includegraphics[height=4cm,width=15cm]{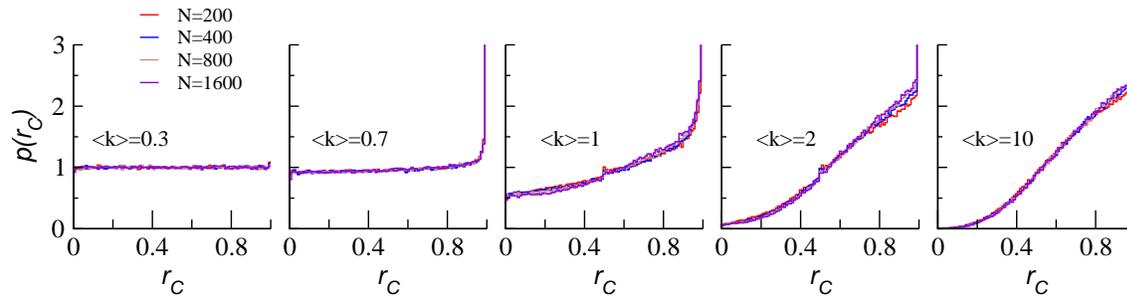}\\
	\vspace{-0.5cm}
	\caption{Histogram of complex eigenvalues spacing ratios $p(r_\mathbb{C})$ from $10^6/2N$ multiplex networks ($A_{NH}$) consisting of directed random networks represented by $A_{dRGE}$ in both layers. The different curves in each panel correspond to different $N$, $N=200, 400, 800, 1600$. Each distribution corresponds to a different average degree $\langle k\rangle$=$0.3$, $0.7$, $1$, $2$, and $10$, represented by vertical dashed lines in Fig.~\ref{fig.1}(d). For a fixed average degree, $p(r_\mathbb{C})$ is invariant with the network size $N$.}
	\label{fig.2}
	\end{center}
\end{figure*}
%%%%%%%%%%%%%%%%%%%%%%%%%%%%%%%%%%%%%%%%%%%%%%%%%%%%%%%%%%%%%

\section{Results and Discussions}
Here, we first present the spectral properties i.e, spacing ratio results of the bi-layer multiplex networks. We compute the average ratio $r_\mathbb{C}$ and $r_\mathbb{R}$ from the eigenvalues obtained from the ensemble of matrices which are numerically diagonalized. For numerical investigations, network size, connection probability, and the average degree are kept the same and are changed simultaneously in both the layers.

\subsection{Diluted Real Ginibre Ensemble} \label{s31}

Fig.~\ref{fig.1} presents the average ratio $r_\mathbb{C}$ as a function of connection probability $p$ for the multiplex networks ($A_{NH}$) with both layers represented by $A_{dRGE}$ and multiplexed by $D_x=1$. For different network sizes, $r_\mathbb{C}$ is computed by taking the average over an ensemble of $10^6/2N$, $2N$ being the size of each multiplex network. Fig.~\ref{fig.1}(a)-(c) depict the dependence of $\langle r_\mathbb{C}\rangle$ on the network size $N$. Fig.~\ref{fig.1}(a) presents the small-size effects for $N<50$, which can be seen in Fig.~\ref{fig.1}(b) plotted for two extreme values of $p$. For $N\leq100$ (Fig.~\ref{fig.1}(b)), as $p$ increases from $p=0$ (when all the nodes are disconnected) to $p=1$ (complete network), $\langle r_\mathbb{C}\rangle$ transits smoothly from the Poisson ensemble statistics $\langle r_\mathbb{C}\rangle \approx0.5$ to the RGE $\langle r_\mathbb{C}\rangle \approx0.73$. Though $\langle r_\mathbb{C}\rangle$ versus $p$ curves follow the same trend for different $N$, transition occurs at the lower values of the $p$ as $N$ increases. This result implies that the PE-RGE transition of $\langle r_\mathbb{C}\rangle$ in a multiplex network begins at lower values of $p$ as $N$ is increased in both layers. Next, for an in-depth understanding of system size dependence on spectral statistics of multiplex networks, the change in position of $\langle r_\mathbb{C}\rangle$-$p$ curves with a change in $N$ is ascertained by defining a parameter $p_o$. $p_o$ is the value of $p$ in both layers of a multiplex network at which $\langle r_\mathbb{C}\rangle$ reaches the midpoint between its smallest and largest values. Fig.~\ref{fig.1}(c) consists of the horizontal dashed line parallel to the $p$ axis marking half of the complete transition at $\langle r_\mathbb{C}\rangle=0.6$. In the inset of Fig.~\ref{fig.1}(d), $p_o$ is plotted for different $N$ values on the log-log scale, which is shown to change linearly with $N$, indicating,

\begin{equation}\label{eq9}
    p_{o}=a(N^\epsilon)
\end{equation}
On fitting the data, we find $\epsilon \approx -1$ and $a\approx1$. Following, we divide $p$ by $p_o$,

\begin{equation}\label{eq10}
    \frac{p}{p_o} \propto \frac{p}{a(N^\epsilon)} \approx \frac{p}{N^{-1}} = Np \equiv \left< k \right> \ ,
\end{equation} 

%%%%%%%%%%%%%%%%%%%%%%%%%%%%%%%%%%%%%%%%%%%%%%%%%%%%%%%%%%%%%%%%%%%%%%%%%%%%%%%%%%
\begin{figure}[t!]
	\begin{center}
		\includegraphics[width=0.5\textwidth]{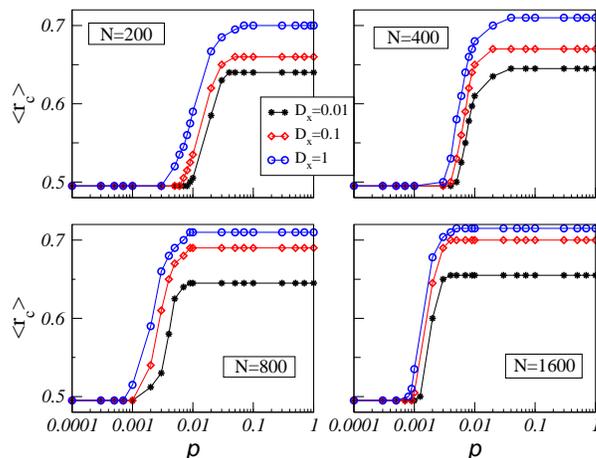}
	\vspace{-0.5cm}
	\caption{Average ratio $\langle r_\mathbb{C}\rangle$ as a function of connection probability $p$ of a multiplex network with both layers represented by diluted real Ginibre ensemble for different values of multiplexing strength $D_x$. Each data point is averaged over $10^6/2N$ random realizations. Decrease in multiplexing strength suppresses the $\langle r_\mathbb{C}\rangle$ transition from PE to RGE.}
	\label{fig.3}
	\end{center}
\end{figure}

\begin{figure}[t!]
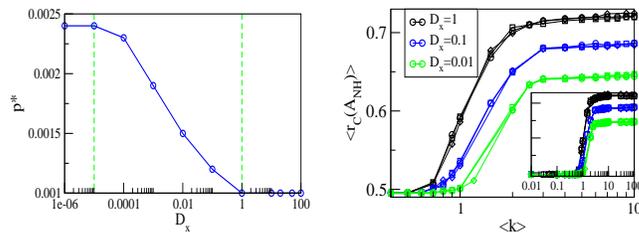

	\begin{center}
	\begin{tabular}{cc}
		\includegraphics[height=3cm,width=4cm]{Dx_pth.eps}&
		\includegraphics[height=3cm,width=4cm]{M1_avg_deg_Dx.eps}\\
	\end{tabular}{}
	\vspace{-0.5cm}
	\caption{ \textbf{Left:}  For a multiplex network consisting of directed random networks represented by $A_{dRGE}$ in both the layers, plot of $p_o$ as a function of $D_x$ for $N=1000$, two vertical lines mark the different regimes of $D_x$. \textbf{Right:} Average ratio $\langle r_\mathbb{C}\rangle$ as a function of average degree $\langle k\rangle$ of a multiplex network with both layers represented by diluted real Ginibre ensemble for different values of multiplexing strength $D_x$. Each $D_x$ value is shown for different network sizes, $N=400, 800$ and $1600$. Each data point is averaged over $10^6/2N$ random realizations.}
	\label{fig.4}
	\end{center}
\end{figure}
%%%%%%%%%%%%%%%%%%%%%%%%%%%%%%%%%%%%%%%%%%%%%%%%%%%%%%%%%%%%%%%%%%%%%%%%%%%%%%%%%%

Fig.~\ref{fig.1}(d) shows $\langle r_\mathbb{C}\rangle$ of a multiplex network as a function of $\langle k\rangle$. The Poisson-Ginibre transition of $r_\mathbb{C}$ by varying $\langle k\rangle$ is observed to transpire on the same $\langle k\rangle$ value for all the network sizes considered here. This indicates that for a fixed average degree, $\langle r_\mathbb{C}\rangle$ of a multiplex network is also fixed independent of the network size $N$. Also, Fig.~\ref{fig.1}(d), $\langle k\rangle<0.7$ marks the region of of PE statistics ($\langle r_\mathbb{C}\rangle \approx0.5$), and for $\langle k\rangle>8$, $\langle r_\mathbb{C}(A_{dRGE})\rangle \approx \langle r_\mathbb{C}$(RGE)$\rangle$. Whereas $0.7<\langle k\rangle<8$ exhibit intermediate statistics as $\langle r_\mathbb{C}\rangle$ transits from PE to RGE statistics in this region. Thus, $\langle k\rangle=0.7$ and $\langle k\rangle=8$ characterize the boundaries between regime of different statistics regions. The result implies that density of connections are determining factor in deciding the behavior of $\langle r_\mathbb{C}\rangle$.

To further substantiate the system size invariance, probability distribution function $p(r_\mathbb{C})$ is presented in Fig.~\ref{fig.2}. For all the five values of $\langle k\rangle$ (marked as vertical lines in Fig.~\ref{fig.1}(d)), the histogram ($p(\langle r_\mathbb{C}\rangle)$) is plotted for various network size of multiplex network, which is observed to fall on a single curve (Fig.~\ref{fig.2}). This result verifies the outcome that $\langle r_\mathbb{C}\rangle$ is independent of system size. Moreover, the results observed here for multiplex networks are aligned with the ones observed in single-layer networks \cite{sl_rat}. However, values of $\langle k\rangle$ that mark the start and end of the PE-Ginibre transition slightly differ in the case of multiplex networks. This behavior is attributed to the additional degree by virtue of the connection between the layers (for instance $D_x=1$ in Fig.~\ref{fig.1} and Fig.~\ref{fig.2}).

%%%%%%%%%%%%%%%%%%%%%%%%%%%%%%%%%%%%%%%%%%%%%%%%%%%%%%%%%%%%%%%%%%%%%%%%%%%%%%%%%%
\begin{figure}[b]
	\begin{center}
		\includegraphics[height=6.5cm,width=8cm]{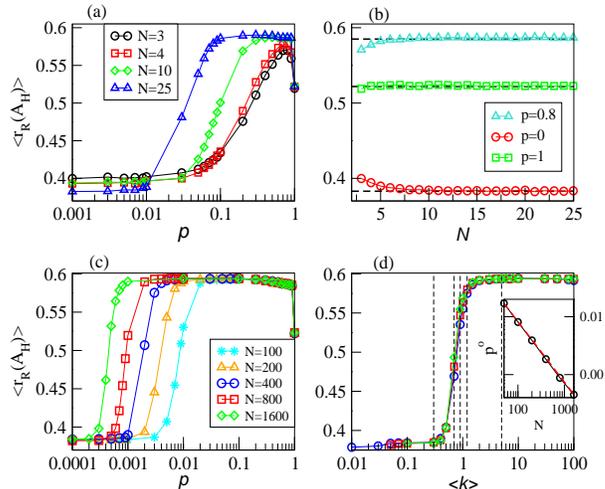}
	\vspace{-0.5cm}
	\caption{(a)-(c) Average spacing ratio $r_\mathbb{R}$ as a function of the probability $p$ of a multiplex networks with both the layers represented by $A_M$, for different network sizes $N$. Horizontal dashed lines (b) for $p$=$0$, $p$=$0.8$ and $p$=$1$ are marked at $r_\mathbb{R}$=$0.38$, $0.6$ and $0.53$ respectively, (c) marked at $r_\mathbb{R}$=$0.49$. (d) $r_\mathbb{R}$ as a function of $\langle k\rangle$ for different network sizes as in (c). In the inset, $p_o$ as a function of $N$ fitted by Eq.~(\ref{eq9}). Vertical dashed lines indicate $\langle k\rangle$=$0.3$, $0.7$, $0.9$, $1.2$, and $5$. Each data point is averaged over $10^6/2N$ random realizations. Increase in network size initiates the transition for lower $p$ whereas once the $\langle k\rangle$ is fixed, $r_\mathbb{R}$ is also fixed. }
	\label{fig.6}
	\end{center}
\end{figure}

\begin{figure}[t!]
	\begin{center}
		\includegraphics[width=1\textwidth]{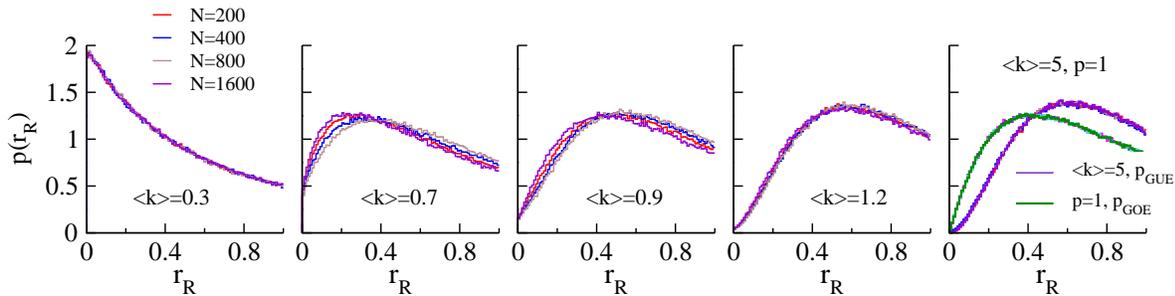}
	\vspace{-0.5cm}
	\caption{Histogram of eigenvalues spacing ratios $p(r_\mathbb{R})$ from $10^6/2N$ multiplex networks consisting of directed random networks represented by $A_{M}$ in both layers. Each distribution corresponds to a different average degree $\langle k\rangle$=$0.3$, $0.7$, $0.9$, $1.2$ and $5$, represented by vertical dashed lines in Fig.~\ref{fig.6}(d). $p(r_\mathbb{R})$ corresponding to each $\langle k\rangle$ is shown for different values of network sizes, $N=200, 400, 800$ and $1600$. For a fixed average degree, $p(r_\mathbb{R})$ is invariant with the network size $N$. Analytic formulae for PE, GUE and GOE in Section.~\ref{s23}. }
	\label{fig.7}
	\end{center}
\end{figure}
%%%%%%%%%%%%%%%%%%%%%%%%%%%%%%%%%%%%%%%%%%%%%%%%%%%%%%%%%%%%%%%%%%%%%%%%%%%%%%%%%%

Next, we focus on inter-layer connections and the influence of $D_x$ on the spectral statistics of multiplex networks. We study the statistics of $\langle r_\mathbb{C}\rangle$ as a function of $p$ and then $\langle k\rangle$ as $D_x$ is varied. Fig.~\ref{fig.3} presents $\langle r_\mathbb{C}\rangle$ as a function of $p$ for different network sizes and the multiplexing strength. We observe that as multiplexing strength is decreased, the $p$ value required for transition to occur increases. Also, for lower values of $D_x$, as $p$ tends to $1$ in both the layers, $\langle r_\mathbb{C}\rangle$ saturates at lower value and do not approach $\langle r_\mathbb{C}($RGE$)\rangle$, hence a partial transition to RGE statistics is observed. This behavior is observed to be the same for all the network sizes. 

The spectral response might be different for different values of $D_x$. To characterize this; we define three regimes of the multiplexing strength $(i)$: weak multiplexing, $D_x<10^{-5}$, where the layers have a negligible interaction between them, $(ii)$: intermediate multiplexing strength, $10^{-5}<D_x<1$, where the layers are coupled, and inter-layer connections play prominent role and $(iii)$: strong multiplexing, $D_x>1$, where layers are superposed with each other.  Fig.~\ref{fig.4} (left) plots $p_o$ (value of $p$ at which $\langle r_\mathbb{C}\rangle$ reaches the midpoint between its smallest and largest values) versus $D_x$ to measure the change in PE-Ginibre transition curves owing to the change in $D_x$ value. It can be seen that in the regimes $(i)$ and $(iii)$, the ratio statistics are not affected by the multiplexing strength, whereas, in regime $(ii)$, $p_o$ decreases as $D_x$ is increased, which reflect that with an increase in $D_x$, the transition occurs with a lower connection probability (Fig.~\ref{fig.4}). Thus, in regime $(ii)$, the  change in $D_x$ affects the PE to RGE transition such that even when both the layers are represented by a complete network at $p=1$, by means of lowering $D_x$ only, the $\langle r_\mathbb{C}\rangle$ is suppressed to transit to the Ginibre ensemble statistics. Additionally, Fig.~\ref{fig.4} (right) presents $\langle r_\mathbb{C}\rangle$ statistics as a function of $\langle k\rangle$. $\langle r_\mathbb{C}\rangle$ versus $k$ curves are observed to follow the same behavior irrespective of the network size as multiplexing strength $D_x$ is changed.

\subsection{Hermitian representation of adjacency matrix} \label{s32}
This section studies the spectral properties of multiplex networks consisting of the directed network represented by Hermitian matrix in both the layers. To analyze the ratio statistics of $A_H$, we make use of $r_R$ (Eq.~(\ref{eq5})). 

Fig.~\ref{fig.6} presents the $\langle r_\mathbb{R} \rangle$ statistics of multiplex networks ($A_H$) consisting of a magnetic adjacency matrix in both layers multiplexed by $D_x=1$. Similar to Fig.~\ref{fig.1} (in the case of $A_{NH}$), here also, we observe small-size effects mainly for $N<50$ (Fig.~\ref{fig.6}(a)), which is further shown in Fig.~\ref{fig.6}(b) for limiting values of connection probability $p$. For $N>100$, $\langle r_\mathbb{R} \rangle$ as a function of $p$ is observed to respond in the same manner as observed for $A_{NH}$. So, we again choose the parameter $p_o$ and plot it with $N$ (inset of Fig.~\ref{fig.6}(d)). After fitting the data and finding the scaling parameter (Eq.~(\ref{eq9})), $r_\mathbb{R}$ is plotted as a function of $\langle k\rangle$ (Fig.~\ref{fig.6}(d)). Again, $\langle r_\mathbb{R}\rangle$ versus $\langle k\rangle$ curves for different network sizes fall on the same curve, demonstrating that for known $\langle k\rangle$ in both the layers, $\langle r_\mathbb{R}\rangle$ of a whole multiplex network can be identified. Thus, if $\langle k\rangle$ is fixed in both the layers, $\langle r_\mathbb{R}\rangle$ of a multiplex network is also fixed.

%%%%%%%%%%%%%%%%%%%%%%%%%%%%%%%%%%%%%%%%%%%%%%%%%%%%%%%%%%%%%%%%%%%%%%%%%%%%%%%%%%
\begin{figure}[t!]
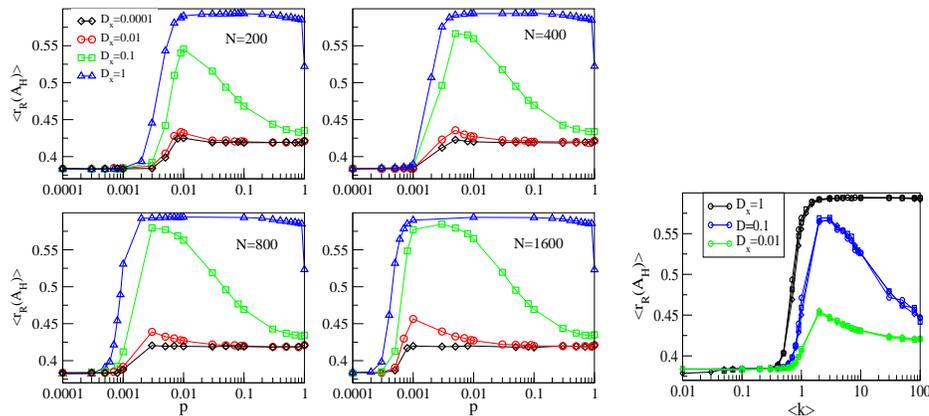

	\begin{center}
	\begin{tabular}{cc}
		\includegraphics[width=0.5\textwidth]{M2_DX.eps}&
		\includegraphics[height=3cm,width=4cm]{M2_avg_deg_Dx.eps}\\
	\end{tabular}{}
	\vspace{-0.5cm}
	\caption{ \textbf{Left:}  Average ratio $\langle r_\mathbb{R}\rangle$ as a function of connection probability $p$ of a multiplex network with both the layers represented by magnetic adjacency matrix $A_M$ for different values of multiplexing strength $D_x$. Each data point is averaged over $10^6/2N$ random realizations. \textbf{Right:} Average ratio $\langle r_\mathbb{R}\rangle$ as a function of average degree $\langle k\rangle$ of a multiplex network with both the layers represented by magnetic adjacency matrix $A_M$ for different values of multiplexing strength $D_x$. Each data point is averaged over $10^6/2N$ random realizations. $\langle r_\mathbb{R}\rangle$ curves corresponding to each $D_x$ is shown for different values of network sizes, $400, 800$ and $1600$. Decrease in multiplexing strength suppresses the $\langle r_\mathbb{R}\rangle$ transition from PE to GUE.}
	\label{fig.8}
	\end{center}
\end{figure}
%%%%%%%%%%%%%%%%%%%%%%%%%%%%%%%%%%%%%%%%%%%%%%%%%%%%%%%%%%%%%%%%%%%%%%%%%%%%%%%%%%

%Though the response of the multiplex network represented by $A_{dRGE}$ ($\langle r_\mathbb{C}\rangle$) is the same as the one represented by $A_M$ ($\langle r_\mathbb{R}\rangle$), few differences are observed. 
The $\langle r_\mathbb{R}\rangle$ vs. $p$ curves show a smooth transition on increasing the $p$ value from $\langle r_\mathbb{R}\rangle \approx0.38$ at $p=0$ to $\langle r_\mathbb{R}\rangle \approx0.6$ at $p\approx0.8$. On further increasing the $p$, $\langle r_\mathbb{R}\rangle$ starts decreasing, and at $p=1$, $\langle r_\mathbb{R}\rangle \approx0.53$ for $N\geq100$ as observed in single layer networks also. 
Here, $\langle r_\mathbb{R}\rangle$ values observed at $p=0$, $p=0.8$ and $p=1$ correspond to RMT ensembles $PE$ ($\approx 0.38$), GUE ($\approx 0.6$) and GOE ($\approx 0.53$), respectively, outlined in references \cite{sl_rat, rat2}. 
Thus, $\langle r_\mathbb{R}\rangle$ of a multiplex network undergoes the triple transition as a function of $p$ from PE to GUE and then GUE to GOE statistics. As $p$ changes, the network in both layers and thus the corresponding adjacency matrix undergoes the transformation; at $p\approx0$, the block matrices $A_1$ and $A_2$ have only real diagonal entries indicating PE statistics, $\langle r_\mathbb{R}\rangle\approx 0.38$. Further increase in $p$ originates random uni-directional links in the network, causing random off-diagonal imaginary entries in the matrix $A_1$ and $A_2$. At this point, GUE like statistics is observed for $A_H$, however, the matrix cannot be said to be alike GUE. As $p\rightarrow1$, all the nodes connect to all other nodes giving rise to bi-directional links and, thus, vanishing imaginary entries in the matrix. Subsequently, GOE like statistics is observed. Note that the results obtained for multiplex networks show similar behavior to single-layer networks \cite{sl_rat}. 

However, the range of $\langle k\rangle$ for which the transition occurs differs slightly from those obtained in single-layer networks, as observed in the case of $A_{NH}$ also. The results imply that on multiplexing two directed random networks, spectral statistics of the whole multiplex network remain the same as that of single-layer networks.
%Also, in the case of $A_M$, $\langle r_\mathbb{R}\rangle$ transition from PE to GUE occurs between $\langle k\rangle=0.5$ to $2$ only, which is narrower as compared to $A_{dRGE}$ where $\langle r_\mathbb{R}\rangle$ transits between $\langle k\rangle=0.8$ to $8$.

Now, we again investigate the impact of multiplexing strength on the relationship between structural properties and spectral properties of the multiplex networks consisting of $A_M$ in both layers. In Fig.~\ref{fig.8}, we present the $\langle r_\mathbb{R}\rangle$ statistics as a function of $p$ for different $D_x$ values and network sizes. First, for lower values of multiplexing strength ($D_x=0.1$ and $0.01$), the connection probability required to begin the transition from PE to GUE statistics and then GUE to GOE statistics increases. Second: for a lowly multiplexed network ($D_x<1$), with an increase in $p$, $\langle r_\mathbb{R}\rangle$ does not complete the transition to GUE/GOE statistics; rather ceases at $\langle r_\mathbb{R}\rangle<\langle r_\mathbb{R}\rangle_{GUE/GOE}$ which shows that the spectral transition can be subdued by decreasing the multiplexing strength ($D_x$). The results, here, infer and propound the multiplexing strength as the regulating parameter for $\langle r_\mathbb{R}\rangle$ statistics as a function of $p$. Also, when $D_x=1$, as p tends to $1$, there is an abrupt transition from GUE to GOE statistics. However, for $D_x<1$, $\langle r_\mathbb{R}\rangle$ transits gradually from near GUE to near GOE statistics. On further decreasing $D_x$ to $0.01$, $\langle r_\mathbb{R}\rangle$ increases slightly with an increase in $p$; however, it remains in near PE statistics regime. Moreover, the behavior of spectral response with the variation in multiplexing strength is observed to be the same for different network sizes considered. Additionally, for a particular $D_x$, $\langle r_\mathbb{R}\rangle$ curves as a function of $\langle k\rangle$ coincide for different $N$ values.

\section{Delocalization Transition} \label{s4}

%%%%%%%%%%%%%%%%%%%%%%%%%%%%%%%%%%%%%%%%%%%%%%%%%%%%%%%%%%%%%%%%%%%%%%%%%%%%%%%%%%
\begin{figure}[b]
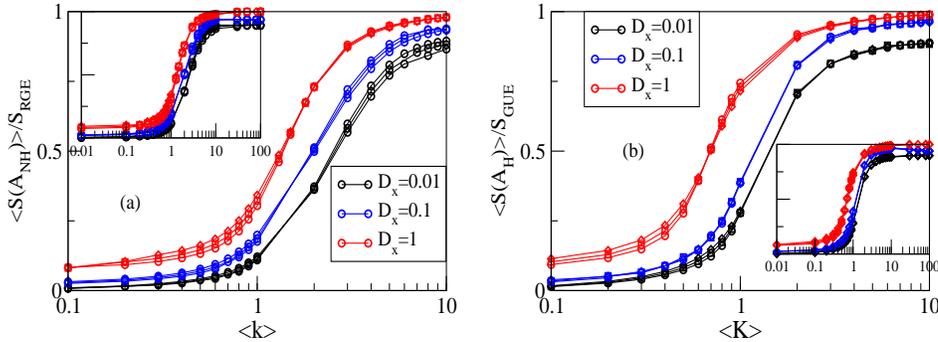

	\begin{center}
	\begin{tabular}{cc}
		\includegraphics[height=4.5cm,width=6cm]{M1_entropy.eps}&
		\includegraphics[height=4.5cm,width=6cm]{M2_Entropy.eps}\\
	\end{tabular}{}
	\vspace{-0.5cm}
	\caption{ \textbf{Left:} Normalized average Shannon entropy $\langle S\rangle$ as a function of the average degree $\langle k\rangle$ for the adjacency matrices of multiplex network with both the layers represented by $A_{dRGE}$ \textbf{Right:} $A_M$ (bottom) corresponding to different network size N=$400$, $800$ and $1600$.  Each data point is averaged over $10^6/2N$ random realizations. Lessening of multiplexing strength restrains the localization to delocalization transition gauged by $\langle S\rangle$.}
	\label{fig.10}
	\end{center}
\end{figure}
%%%%%%%%%%%%%%%%%%%%%%%%%%%%%%%%%%%%%%%%%%%%%%%%%%%%%%%%%%%%%%%%%%%%%%%%%%%%%%%%%%

Entropy has been widely studied as a measure of the degree of irregularity \cite{ent_irr}, disorder \cite{ent_dis}, complexity \cite{ent_cmplx}, and randomness \cite{ent_ran} in various systems. Also, it has been used to study the localization-delocalization transition in several systems \cite{l_del,l_del1}. In this section, we apply entropy analysis, first: to quantify localization-delocalization transition in multiplex networks consisting of directed networks in both layers. Secondly, to verify the PE-RGE transition of $A_{NH}$ and PE-GUE transition of $A_H$ by a change in the corresponding eigenvectors from the localized regime (PE) to extended (RGE or GUE). In the case of multilayer networks, delocalization transition has been studied using different measures \cite{MN_loc, MN_loc1, MN_loc2}. Here, we make use of Shannon entropy $S$, which for eigenvector $\psi^i$ corresponding to $i_{th}$ eigenvalue ($\lambda_i$) is defined as
\begin{equation}
S^i = -\sum_{j = 1}^n \mid \Psi_j^i \mid^2 \ln \mid \Psi_j^i \mid^2 
\label{S}
\end{equation}

$S$ has been used as an exemplary index for eigenvector localization-delocalization transition \cite{ent_loc} as it provides information about the number of principal components of the eigenvector $\psi^i$. $\langle S\rangle$ has also been used to calculate the localization length in several systems \cite{loc_len}. Here, to compute Shannon entropy $\langle S\rangle$ of multiplex networks, we numerically diagonalize the corresponding matrices of a large ensemble. Then the average is taken over all eigenvectors such that for an ensemble size of $10^6/2N$, $\langle S\rangle$ is computed over $10^6$ eigenvectors.
In the case of single-layer networks, for both $A_{dRGE}$ and $A_M$ \cite{sl_rat}, when p=$0$, only diagonal entries are present in the matrix, due to which, in the corresponding eigenvector, there is only one component with a magnitude$\approx1$, resulting in $S\approx0$. As $p$ is increased to $1$, now, because of the presence of off-diagonal entries, the corresponding eigenvectors will have nearly equal components resulting in $S\approx \ln(n)$. However, in the case of multiplex networks, an additional entry is present in each row of the corresponding adjacency matrix ($A_{NH}/A_H$), owing to interlayer connection ($D_x=1$) resulting in $\langle S\rangle \neq0$ even at $p=0$. 

In Fig.~\ref{fig.10}, $\langle S\rangle$ versus $\langle k\rangle$ is presented for different network sizes ($N$) and multiplexing strength ($D_x$) of multiplex networks. Fig.~\ref{fig.10}(a) and \ref{fig.10}(b) show $\langle S\rangle$ for multiplex networks represented by $A_{NH}$ and $A_H$, respectively. For a particular multiplexing strength, as $p$ increases from $0$ to $1$, $\langle S\rangle$ is observed to manifest a smooth transition from $\langle S\rangle$=$0$ (when only diagonal entries are present, and no nodes are connected) to its maximum value $S_{max}$ (when all the nodes are connected). Also, for different network sizes $N$, $\langle S\rangle$ shows the same behavior with respect to $\langle k\rangle$. Now, when $p$=$0$ in both the layers, eigenvectors of a multiplex network are localized as $\langle S\rangle\approx0$, and as $p$ tends to $1$, $\langle S\rangle$ reaches its maximum value, then the eigenvectors are delocalized. The results obtained are the same for both the multiplex networks consisting of networks corresponding to $A_{dRGE}$ and $A_M$ in both layers. However, $S_{max}$ for $A_{NH}$ and $A_{M}$ is different, so to get the normalized value of average entropy, $\langle S\rangle/S_{max}$ is computed for all the values of $\langle k\rangle$. Also, subject to $A_{NH}$, $\langle S\rangle$ is $\approx S_{RGE}$ which is the Shannon entropy of eigenvectors of RGE matrices. Note that $S_{RGE}$ considered here is referred from the recent work \cite{sl_rat} where it was computed numerically for different network sizes. In the case of a multiplex network $A_H$, $S_{Max}$=$S_{GUE}$ and $S_{GUE}\approx \ln(n/1.53)$ \cite{ent_loc}. Now, as we decrease the multiplexing strength, the average degree required to begin the transition increases. Also, for lower $D_x$, as $\langle k\rangle$ is increased, $\langle S\rangle$ is observed to saturate at lower values than $S_{max}$.
Moreover, for $D_x$=$1$, because of the additional inter-layer connection, even at $p$=$0$, $\langle S\rangle>0$ whereas for $D_x$=$0.1$ and $0.01$, $\langle S\rangle \approx0$. 
%The fact that for $\langle k\rangle<1$, $\langle S\rangle \neq 0$ indicates that even when the network has most of the nodes isolated, few eigenvector entries are comparatively higher than the others. 
%Thus, in Fig.~\ref{fig.10}, $\langle S\rangle$ versus $\langle k\rangle$ curves show the localization-to-delocalization transition of the eigenvectors of the adjacency matrix of multiplex networks with $A_{dRGE}$ (Fig.~\ref{fig.10}(a)) and $A_M$ (Fig.~\ref{fig.10}(b)) depicting both layers.

\section{Conclusion} \label{s5}
To conclude, we have numerically analyzed the spectral properties of multiplex networks consisting of directed random networks in both layers. In particular, by employing the spacing ratio measure, we investigated the average ratio as a function of various structural parameters of the network in layers. Without a requisition of the unfolding, the eigenvalue spacing ratio can determine the spectral statistics of various systems. Here, we study ER random networks by characterizing them in two different manners, ($i$) diluted real Ginibre ensemble (dRGE), which are sparse non-Hermitian random matrices and are analyzed using the complex spacing ratio ($r_\mathbb{C}$) (Eq.~(\ref{eq2})), ($ii$) sparse Hermitian matrices for which real spacing ratio ($r_\mathbb{C}$) is computed (Eq.~(\ref{eq3})). Multiplexing two networks yields the spacing ratio statistics, as that of the corresponding single-layer networks \cite{sl_rat}. The average ratio $\langle r_\mathbb{C}\rangle$/$\langle r_\mathbb{R}\rangle$ is shown to transit smoothly from PE to RGE/GUE statistics as a function of $p$. The $p$ value for which this transition occurs manifests a decrease with an increase in the network size. After scaling and varying $\langle r\rangle$ as a function of $\langle k\rangle$, the transition curves for different network size are observed to coincide on a single universal curve. 
Further, in recent years, studies on multiplex networks have revealed additional information than those of the corresponding single-layer networks. To illustrate this, we analyzed the spacing ratio statistics for different multiplexing strengths. For a multiplex network represented by a diluted version of real Ginibre ensemble matrices in both layers, varying the multiplexing strength affect the ratio statistics in two ways. First, at $D_x$=$1$, $\langle r_\mathbb{C}\rangle$ shows a transition from PE to RGE statistics as $p$ tends to $1$. However, for $D_x<1$, as $p$ tends to $1$, $\langle r_\mathbb{C}\rangle$ saturates at lower values, and a complete transition to RGE statistics is suppressed. Second,  with a decrease in the multiplexing strength, more connection probability is required to commence the transition. Next, for multiplex networks consisting of sparse Hermitian matrices in both the layers, a triple transition PE-GUE-GOE is observed with an increase in the connection probability. We show that upon decreasing the multiplexing strength hinders the PE-GUE and GUE-GOE statistics transition. Also, for $D_x$=$1$, as $p$ tends to $1$, a sudden transition occurs from GUE-GOE statistics which for lower values of $D_x$ is found to occur gradually.\\
We wish to note here that the results presented here are akin to the phase transitions concerning Raman scattering \cite{mat1} and $2D$ materials such as MoS$_{2}$ nanoflakes \cite{mat2}. The results achieved here using multilayer model can be helpful in better understanding of several underlying mechanisms and dynamics of such systems. Additionally, we analyze the localization-delocalization transition of the eigenvectors of multiplex networks using Shannon entropy. We show that by decreasing the multiplexing strength, even for $p$=$1$ (when the network is complete), the eigenvectors of a multiplex network are not wholly delocalized, which again implies that $D_x$ can act as a suppressing parameter in studying several processes in multiplex networks. 
In recent years, eigenvalue spacing ratio statistics has gained enormous attention from researchers because of its wide application in various fields such as condensed matter, chaotic systems, etc. Further, spacing ratio technique has already been used to measure the localization-delocalization transition in different systems, including single-layer networks. We believe that spacing ratio technique can potentially be used to study various structural and dynamical properties of complex systems represented by multiplex networks having directed connections. For instance, eigenvalues of laplacian matrix can give insight into synchronization of corresponding complex network. This work has focused on adjacency matrix only, one can analyze the laplacian of multiplex directed networks.

\ack 
SJ and Tanu acknowledge Govt. of India, BRNS Grant No. 37(3)/14/11/2018-BRNS/37131 for financial support and SRF fellowship, respectively. SJ thankfully acknowledges DST grant SPF/2021/000136. We thank Ranveer Singh and Stefano Boccaletti for useful suggestions and interesting discussions.  

\section*{References}
%\bibliographystyle{iopart-num}
%\bibliography{references.bib}
\providecommand{\newblock}{}

\end{document}